\pdfoutput=1

\documentclass[aps,twocolumn,showpacs,floatfix,superscriptaddress]{revtex4-1}

\usepackage{dcolumn}
\usepackage{ulem}
\usepackage{makecell}
\usepackage[utf8]{inputenc} 
\usepackage{graphicx}

\usepackage{amsfonts}
\usepackage{amsmath,bm,amssymb}
\mathchardef\mhyphen="2D

\usepackage[usenames,dvipsnames]{color}
\usepackage{kotex}

\usepackage{comment}

\usepackage{usebib} 

\begin{document}

\title{Microscopic understanding of magnetic interactions in bilayer CrI$_3$}

\author{Seung Woo Jang}
\affiliation{Department of Physics, Korea Advanced Institute of Science and Technology (KAIST), Daejeon 34141, Republic of Korea}
\author{Min Yong Jeong}
\affiliation{Department of Physics, Korea Advanced Institute of Science and Technology (KAIST), Daejeon 34141, Republic of Korea}
\author{Hongkee Yoon}
\affiliation{Department of Physics, Korea Advanced Institute of Science and Technology (KAIST), Daejeon 34141, Republic of Korea}
\author{Siheon Ryee}
\affiliation{Department of Physics, Korea Advanced Institute of Science and Technology (KAIST), Daejeon 34141, Republic of Korea}
\author{Myung Joon Han}
\affiliation{Department of Physics, Korea Advanced Institute of Science and Technology (KAIST), Daejeon 34141, Republic of Korea}
\email{mj.han@kaist.ac.kr}

\begin{abstract}
We performed a detailed microscopic analysis of the inter-layer magnetic couplings for bilayer CrI$_3$. As the first step toward understanding 
the recent experimental observations and utilizing them for device applications, we estimated magnetic force response as well as total energy. Various van der Waals functionals unequivocally point to the ferromagnetic ground state for the low-temperature structured bilayer CrI$_3$ which is further confirmed independently by magnetic force response calculations. The calculated orbital-dependent magnetic forces clearly show that $e_g$-$t_{2g}$ interaction is the key to stabilize this ferromagnetic order. By suppressing this ferromagnetic interaction and enhancing antiferromagnetic orbital channels of $e_g$-$e_g$ and $t_{2g}$-$t_{2g}$, one can realize the desirable antiferromagnetic order. We showed that high-temperature monoclinic stacking can be the case. Our results provide unique information and insight to understand the magnetism of multi-layer CrI$_3$ paving the way to utilize it for applications. 
\end{abstract}

\maketitle


Recently, magnetism in 2-dimensional (2D) van der Waals (vdW) materials has attracted great attention \cite{wang_raman_2016,tian_magneto-elastic_2016,lee_ising-type_2016,gong_discovery_2017,bonilla_strong_2018,kim_charge-spin_2018,wang_electric-field_2018,fei_two-dimensional_2018, huang_layer-dependent_2017,zhong_van_2017,seyler_ligand-field_2018, jiang_electric-field_2018,jiang_controlling_2018,huang_electrical_2018,song_giant_2018,klein_probing_2018,wang_very_2018,kim_one_2018,song_voltage_2019,cardoso_van_2018}. It is not just due to their novelty from a fundamental physics point of view, but also to their great potential for device applications. Importantly, however, understanding the microscopic nature of those magnetic interactions is far from complete, posing an outstanding challenge for theoretical material science. The difficulty is partly attributed to that there is no well-established exchange-correlation functional for describing vdW interaction. Although there are several promising functionals now available \cite{grimme_semiempirical_2006,grimme_consistent_2010,grimme_effect_2011,tkatchenko_accurate_2009,tkatchenko_accurate_2012,ambrosetti_long-range_2014,steinmann_comprehensive_2011,steinmann_generalized-gradient_2011,dion_van_2004,roman-perez_efficient_2009,klimes_van_2011,lee_higher-accuracy_2010}, the reliable description of this weak interaction is still quite challenging for first-principles simulations. Another difficulty is related to the lack of conventional physical `picture' to  describe magnetic interactions in these materials such as superexchange and double-exchange model for typical ionic solids. Without such an intuitive picture a clear understanding of the observed phenomena and utilizing them for device application are severely hampered.

An outstanding example to demonstrate this type of challenge is CrI${_3}$. Just after its first realization of CrI${_3}$ monolayer \cite{huang_layer-dependent_2017}, this magnetic insulating 2D material has generated great research interest  \cite{huang_layer-dependent_2017,zhong_van_2017,seyler_ligand-field_2018, jiang_electric-field_2018,jiang_controlling_2018,huang_electrical_2018,song_giant_2018,klein_probing_2018,wang_very_2018,kim_one_2018,song_voltage_2019,cardoso_van_2018}. While the monolayer ferromagnetism is well reproduced by first-principles calculations \cite{wang_electronic_2011,sivadas_magnetic_2015,jiang_spin_2018,lado_origin_2017,zheng_tunable_2018}, understanding the multi-layer systems remains quite elusive. A recent magneto-optical
Kerr effect (MOKE) measurement of bilayer CrI${_3}$ shows the vanishing Kerr rotation which indicates the inter-layer antiferromagnetic (AFM) coupling by excluding ferromagnetism \cite{huang_layer-dependent_2017}. AFM order is further confirmed by magneto-photoluminescence (PL) \cite{seyler_ligand-field_2018}, reflective magnetic
circular dichroism (RMCD), and transport measurements \cite{jiang_electric-field_2018,song_giant_2018,klein_probing_2018,huang_electrical_2018,kim_one_2018}. On the contrary, however, the first-principles calculations report that the ferromagnetic (FM) spin order with low-temperature (LT) structure is energetically most favorable \cite{jiang_stacking_2018,soriano_interplay_2018}.

In this work, we investigate the magnetic interactions of bilayer CrI${_3}$. First, we performed total energy calculations with various forms of exchange-correlation functionals supported by the state-of-the-art cRPA (constrained random phase approximation) technique. It is found that FM inter-layer coupling is always favorable in LT stacking, which provides a stronger indication of FM order in this structure. Further, we performed the magnetic force response calculation which can directly measure the spin-spin interaction independent of total energy values. The calculated magnetic responses unequivocally point to the FM inter-layer coupling which is another strong evidence. In order to unveil the microscopic origin of inter-layer couplings, 
we investigate the orbitally-decomposed magnetic interactions using our recent implementation \cite{yoon_reliability_2018}. Surprisingly, Cr-$e_g$ orbitals are found to play an important role. Our calculations clearly show that the second-neighbor $e_g$-$t_{2g}$ interactions are the main source of FM order in LT-structured bilayer CrI$_3$. This coupling is significantly suppressed and becomes comparable with AFM $e_g$-$e_g$ interaction in high-temperature (HT) structure. Our analysis provides the detailed information and insight which pave the way toward understanding the fascinating phenomena reported in this material \cite{huang_layer-dependent_2017,seyler_ligand-field_2018} and utilizing them for device applications \cite{jiang_electric-field_2018,jiang_controlling_2018,huang_electrical_2018,song_giant_2018,klein_probing_2018,wang_very_2018,kim_one_2018,song_voltage_2019}.

\begin{figure*}[!t]
	\begin{center}
		\includegraphics[width=0.75\textwidth,angle=0]{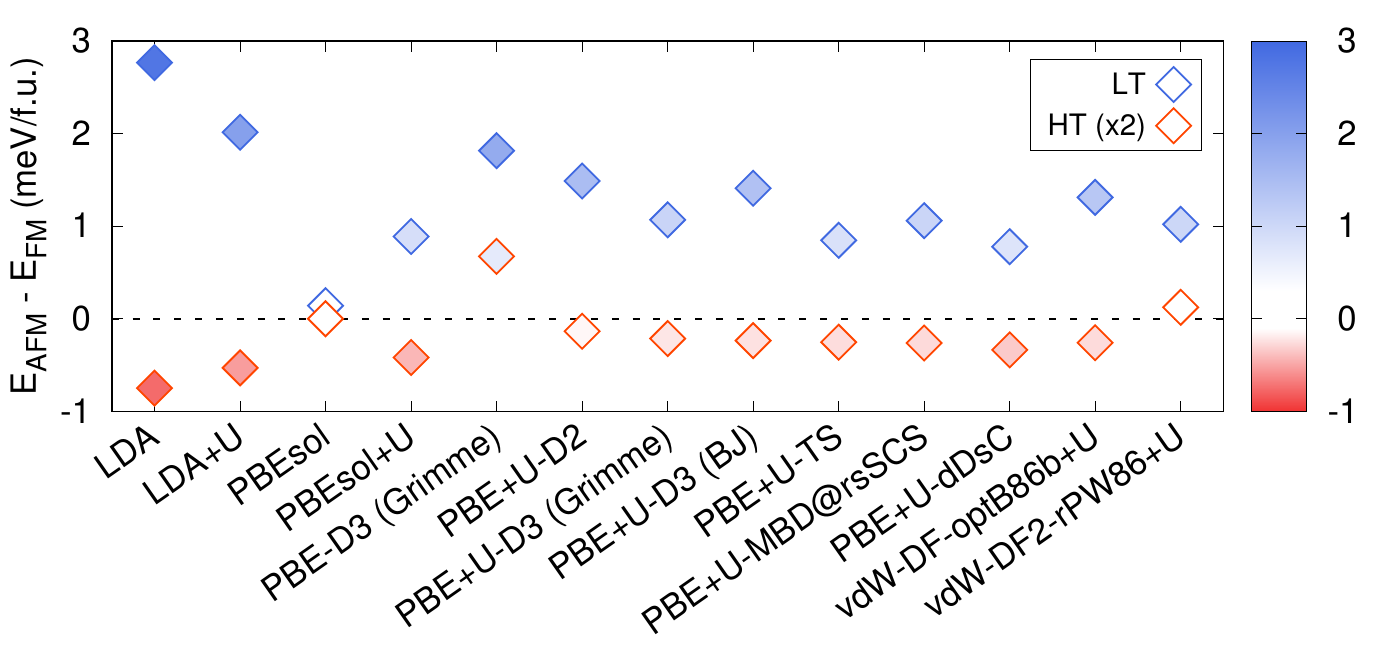}
		\caption{The calculated total energy difference between the layered AFM and FM with various functionals. $E_{\textrm{FM}}$ and $E_{\textrm{AFM}}$ denote the calculated total energy for the inter-layer FM and AFM phase, respectively. The blue and red symbols represent the results of LT and HT stacking, respectively (see the inset). The filling colors represent the calculated value of energy difference (see the color bar). For a clear presentation, the calculated values for HT stacking have been multiplied by 2.
			\label{Figure 1}}
	\end{center}
\end{figure*}


The total energy calculations with different vdW functionals were carried out with VASP code \cite{vasp}. We considered many different exchange-correlation functionals including  LDA (local density approximation) \cite{ceperley_ground_1980,perdew_self-interaction_1981}, PBE \cite{PBE}, PBEsol \cite{perdew_restoring_2008}, D2  \cite{grimme_semiempirical_2006}, D3 (Grimme), D3 (BJ) \cite{grimme_consistent_2010,grimme_effect_2011}, TS \cite{tkatchenko_accurate_2009}, MBD$@$rsSCS \cite{tkatchenko_accurate_2012,ambrosetti_long-range_2014}, dDsC \cite{steinmann_comprehensive_2011,steinmann_generalized-gradient_2011}, vdW-DF-optB86b \cite{dion_van_2004,roman-perez_efficient_2009,klimes_van_2011}, and vdW-DF2-rPW86 \cite{lee_higher-accuracy_2010}. The 600~eV energy cutoff and 9 $\times$ 9 $\times$ 1 {\bf k}-points were used for the first Brillouin zone. We found this numerical settings are enough to achieve the desirable accuracy. Atomic positions were relaxed with the force criterion of 1~meV/{\AA} and the lattice constants fixed to the experimental values; $a$=$b$=6.867~{\AA} for rhombohedral LT and $a$=$b$=6.863~{\AA} for monoclinic HT stackings \cite{mcguire_coupling_2015}. We also found that the magnetic ground state is not changed when using the optimized lattice constants. We took the vacuum distance of $\sim$20~{\AA} which is found to be large enough to simulate the experimental situation.
For DFT$+U$ (density functional theory plus $U$) method \cite{LDAU_review}, we used so-called FLL (fully localized limit) version of DFT$+U$ based on charge density \cite{Anisimov_93,Ryee}. It is found that spin-orbit coupling does not change the magnetic ground state which is consistent with the previous calculation by Sivadas {\it et al} \cite{sivadas_stacking-dependent_2018}. For magnetic force theory (MFT) calculation \cite{MFT1,MFT2,han_electronic_2004,yoon_reliability_2018}, we used OpenMX software package \cite{openmx,LCPAO} which is based on LCPAO (linear combination of pseudoatomic orbitals) formalism. 8 $\times$ 8 $\times$ 1 {\bf k} mesh was used for MFT calculation. The D3 method of Grimme {\it et al.} was used for the vdW correction \cite{grimme_consistent_2010} in this process since it best reproduces the lattice constants and the cell volume for bulk CrI$_3$. For the estimation of interaction parameters, the constrained random phase approximation (cRPA) \cite{cRPA,Bluegel} was performed with Ecalj package \cite{ecalj}. We used so-called $d$-$dp$ model \cite{miyake_d-_2008,Vaugier} as derived by the maximally localized Wannier function technique \cite{MLWF}.


While many of theoretical studies have been devoted to bulk and monolayer CrI$_3$ \cite{wang_electronic_2011,sivadas_magnetic_2015, zhang_robust_2015,mcguire_coupling_2015,jiang_spin_2018,liu_exfoliating_2016,zheng_tunable_2018,larson_raman_2018-1,webster_distinct_2018,lado_origin_2017,djurdjic-mijin_lattice_2018}, the inter-layer interaction of bilayer or multi-layer is largely unexplored. Three first-principles investigations have been reported quite recently which focus on the stacking patterns to understand the inter-layer coupling \cite{wang_very_2018, jiang_stacking_2018, soriano_interplay_2018}. The total energy calculations based on GGA (generalized gradient approximation) or GGA$+U$ with a certain type of vdW correction show that the FM inter-layer spin order with LT (rhombohedral) structure is energetically most favorable for bilayer CrI$_3$ in contrast to experimental observations \cite{wang_very_2018, jiang_stacking_2018, soriano_interplay_2018}.

Let us start by noting that the computation studies are limited to a couple of vdW functional recipes, namely, (so-called) `svdW-DF' \cite{wang_very_2018}, `vdW-DF-optB86b$+U$' \cite{jiang_stacking_2018}, and `PBE-D2' \cite{soriano_interplay_2018}. Since the universal vdW functional within density functional framework is not well established yet, it is strongly required to confirm whether this is a physically reasonable solution, not an artifact, especially considering the inconsistency with experiments. Further, this material CrI$_3$ is known to be a Mott insulator \cite{mcguire_coupling_2015} for which the conventional LDA or GGA functional does not give the reasonable electronic structure. This is the reason for several recent studies to adopt LDA/GGA$+U$ functionals \cite{liu_exfoliating_2016,olsen_assessing_2017,jiang_stacking_2018}. While DFT$+U$ is certainly the better choice for Mott insulators, its final result critically depends on the choice of `interaction parameters' such as Hubbard $U$ and Hund $J_H$. Indeed, the previous study by Jiang {\it et al.} shows that the spin ground state of HT-phase bilayer CrI$_3$ changes from FM to AFM at around $U$=2.5 eV \cite{jiang_stacking_2018}.

\begin{figure*}[!t]
	\begin{center}
		\includegraphics[width=0.8\textwidth,angle=0]{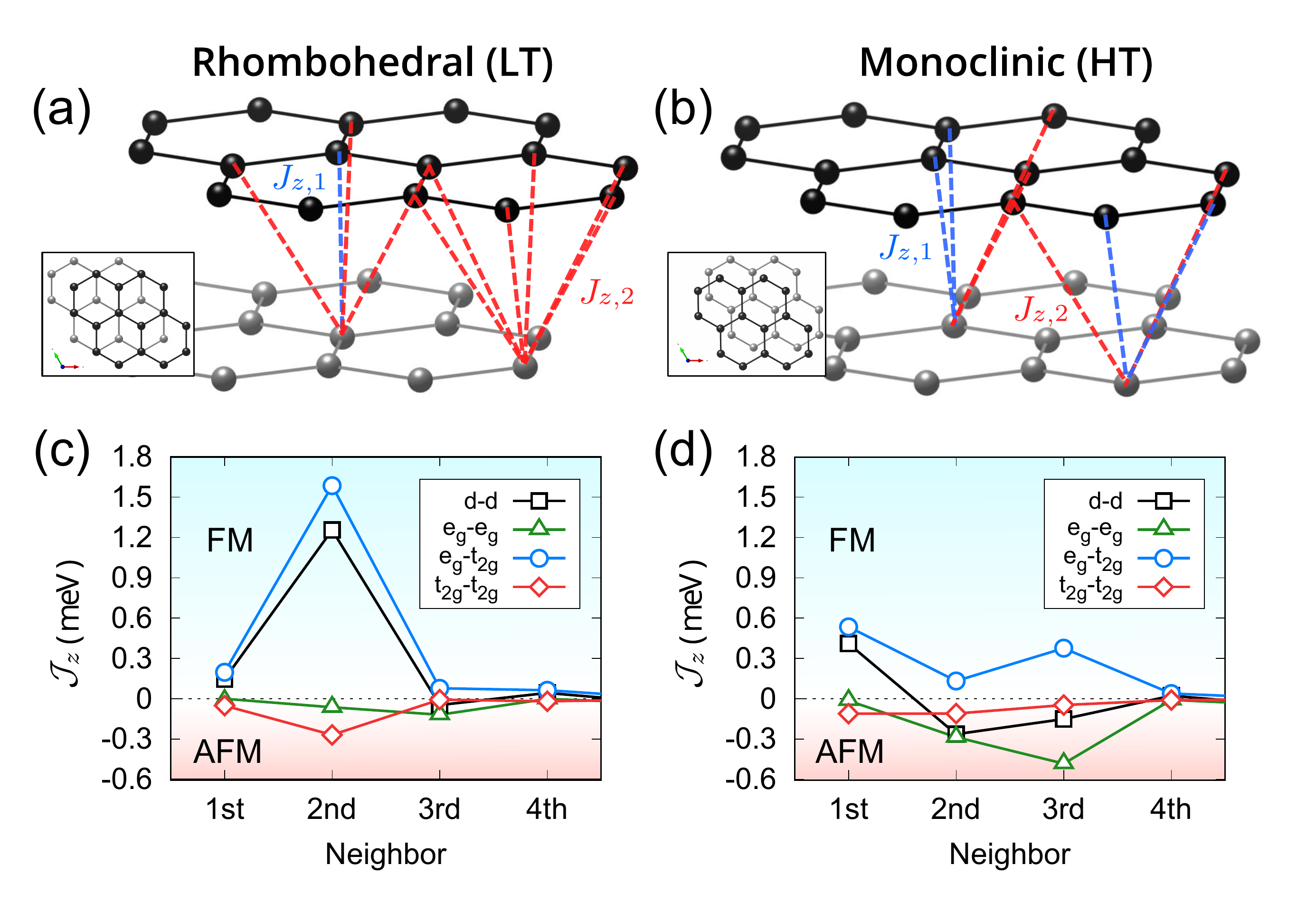}
		\caption{(a, b) The schematic picture of inter-layer magnetic interactions in bilayer CrI$_3$ of (a) LT rhombohedral and (b) HT monoclinic structure. The first ($J_{z,1}$) and the second neighboring inter-layer couplings ($J_{z,2}$) are represented by blue and red colors, respectively. The black and gray atoms represent Cr.  Insets of (a) and (b) represent the top view of each stacking. (c, d) The calculated $\mathcal{J}_z$ values as a function of neighboring distance. The positive and negative values correspond to FM and AFM interaction, respectively. The interactions through each orbital pair channel are represented by different colors and symbols as denoted in the inset of (d). The black lines with squares refer to the total $d$ orbital interactions which correspond to the sum of $e_g$-$e_g$, $e_g$-$t_{2g}$, and $t_{2g}$-$t_{2g}$ interactions.
			\label{Figure 2}}
	\end{center}
\end{figure*}

Here we first estimate the interaction parameters based on the most advanced scheme, namely, cRPA \cite{cRPA,Bluegel} which is computationally demanding but known as quite reliable \cite{wehling_strength_2011,werner_satellites_2012,yamaji_first-principles_2014,jang_direct_2016,jang_charge_2018}. The calculated on-site Coulomb repulsion $U=2.0$ eV for the bulk CrI$_3$ and $U=2.9$ eV for monolayer. The Hund interaction is found to be $J_H=0.7$ eV for both bulk and monolayer. It is noted that the on-site electron correlation $U$ is significantly enhanced by $\sim$30\% when the system dimension is reduced. This value is used for our bilayer calculations.

Now we investigate the total energy profile to confirm the magnetic ground state of LT structure. We consider most of the available vdW functionals including eight different correction types. The results are summarized in Figure~\ref{Figure 1}. It is clearly shown that the inter-layer AFM order is not stabilized in LT stacking; see the blue diamonds in Figure~\ref{Figure 1}. For all of the functional choices, the calculated total energies of AFM order are larger than those of FM by more than $\sim$0.78 meV per formula unit except for PBEsol (see Figure~\ref{Figure 1}). Our result is a strong evidence that the ground state of bilayer CrI$_3$ is FM in the LT stacking.

As discussed in the previous studies, the AFM inter-layer coupling is important for device application \cite{seyler_ligand-field_2018, jiang_electric-field_2018,jiang_controlling_2018,song_giant_2018,klein_probing_2018,wang_very_2018,kim_one_2018,song_voltage_2019,huang_electrical_2018}. To explore this possibility and to understand the fascinating recent experimental observations such as voltage-controlled magnetism \cite{jiang_electric-field_2018,jiang_controlling_2018,huang_electrical_2018} and GMR (giant magnetoresistance) \cite{wang_very_2018,kim_one_2018,song_giant_2018,klein_probing_2018,song_voltage_2019}, the key first step is to have the microscopic picture of interlayer interactions. Here we note that the conventional interaction model such as superexchange is not relevant to this case of vdW materials as the second order hopping does not connect even the first neighboring Cr sites. Thus, the simple-model-based approach can hardly be successful, and the first-principles-based simulation is desirable. With this motivation, we performed the MFT calculations in which the magnetic interaction, $J$, is calculated as a response to the small angle tilting of spin rotations  \cite{MFT1,MFT2,han_electronic_2004,yoon_reliability_2018}.

\begin{figure*}[!t]
	\begin{center}
		\includegraphics[width=0.75\textwidth,angle=0]{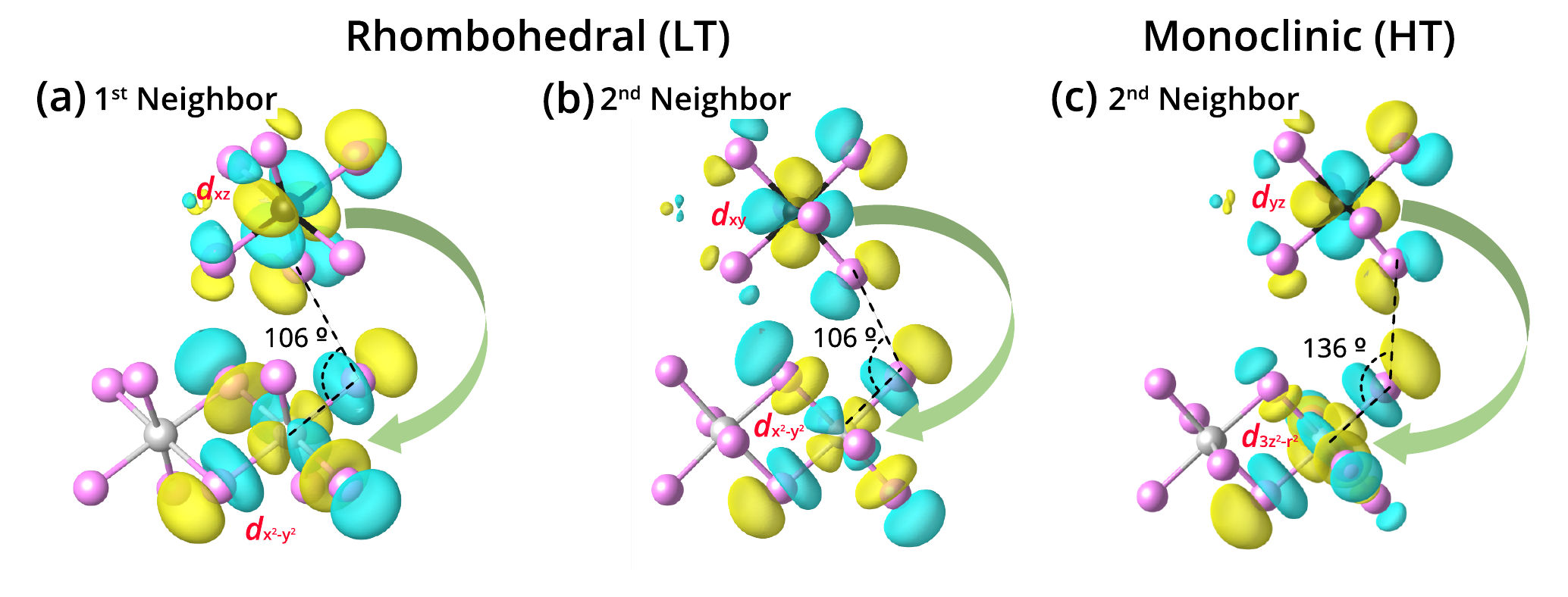}
		\caption{The maximally localized Wannier orbitals for the (a)-(b) LT and (c) HT stacking. The green arrows highlight the $e_g$-$t_{2g}$ magnetic interaction for the first neighbor in (a) and the second neighbor pairs in (b)-(c).}
		\label{Figure 3}
	\end{center}
\end{figure*}

Our results of MFT are summarized in Figure~\ref{Figure 2} where  the $n$-th neighbor out-of-plane interaction $\mathcal{J}_{z,n}$ is defined as the sum of all pairs of $J_{z,n}$ reflecting the corresponding coordination number. For the LT-stacked bilayer CrI$_3$, the calculated interlayer coupling $\mathcal{J}_z$ is FM; see the black line with squares in Figure~\ref{Figure 2}c. The nearest-neighbor $\mathcal{J}_{z,1}$ and the second neighbor $\mathcal{J}_{z,2}$ are both FM\cite{comment-1} whereas the longer-range inter-layer interactions ($\mathcal{J}_{z,n\geq 3}$) are negligibly small. This is another meaningful confirmation that the AFM order is not stabilized in LT stacking. It is important to note that MFT calculation does not rely on total energy information, but just utilizes eigenfunctions and eigenvalues \cite{MFT1,MFT2,han_electronic_2004,yoon_reliability_2018}. Indeed, when we calculated $\mathcal{J}_z$ values based on the AFM solution of LT structure (which is not the ground state), the response function favors the spin flip, indicative of the FM ground state. Our MFT results provide an independent additional confirmation for FM interlayer coupling in LT structure.

In order to have further insights, we calculate orbitally-decomposed magnetic interactions, which is a unique and useful feature of MFT. As shown in Figure~\ref{Figure 2}c, the dominant contribution comes from FM $e_g$-$t_{2g}$ channels (see the blue line with circles). The $e_g$-$e_{g}$ and $t_{2g}$-$t_{2g}$ orbital interactions are AFM but significantly weaker. This detailed microscopic information provides the unique insights to understand the magnetism of this material. It is remarkable that $e_g$ orbitals play the important role which should be magnetically inactive in the pure ionic picture of Cr$^{3+}$. Due to the hybridization with I-$p$ orbitals, the $e_g$ states carry the sizable moments as shown in Table~1. This feature has not been properly recognized before \cite{jiang_stacking_2018,soriano_interplay_2018} and it demonstrates an intriguing nature of vdW magnetic materials distinctive from the typical ionic Mott insulators.

\begin{table}[b]
	\caption{The calculated orbital-resolved electron occupations and magnetic moments for HT stacking. The $\uparrow$ and $\downarrow$ denote the up and down spins, respectively. The $e_g$ and $t_{2g}$ states are defined in each atomic local axis.}
	\label{Table 1}
	\begin{tabular}{lllll}
		\hline
		&        & $N_{\uparrow}$ & $N_{\downarrow}$  & $M$    \\
		\hline
		& $e_g$    & 1.10            & 0.51 & 0.59   \\
		Cr & $t_{2g}$ & 2.85            & 0.15 & 2.70   \\
		& $d$      & 3.95            & 0.66 & 3.29   \\
		I  & $p$      & 2.61            & 2.75 & $-$0.14\\
		\hline
	\end{tabular}
\end{table}

One important implication of our orbital-decomposed $\mathcal{J}$ results is that, if one can suppress $e_g$-$t_{2g}$ interaction and enhance $e_g$-$e_{g}$ and/or $t_{2g}$-$t_{2g}$, AFM order can be stabilized which is desirable for many purposes \cite{jiang_electric-field_2018,jiang_controlling_2018,huang_electrical_2018,song_giant_2018,klein_probing_2018,wang_very_2018,kim_one_2018,song_voltage_2019}. As one example for this, we calculated the HT-stacked monoclinic bilayer (see Figure~\ref{Figure 2}b), and the results are presented in Figure~\ref{Figure 2}d. The HT structure of bulk CrI$_3$ is specified by the space group C2/m and the LT structure by R$\bar{3}$. Importantly, the different stacking leads to the different number of magnetic couplings. For bilayer CrI$_3$, the HT stacking has four first-neighbor and four second-neighbor couplings whereas the LT stacking has one nearest and nine next-nearest neighbors. Due to the change of hopping routes (to be discussed further below) as well as the different numbers of neighboring sites, the orbital interaction profile is notably different from that of LT structure. While the 
$e_g$-$t_{2g}$ interactions $\mathcal{J}^{e_{g}\textrm{-}t_{2g}}_{z}$ are still FM, the second neighbor $\mathcal{J}^{e_{g}\textrm{-}t_{2g}}_{z,2}$ becomes significantly weaker. As a result, the total $\mathcal{J}_{z,2}$ becomes AFM in HT structure.
It is also noted that the third neighbor $\mathcal{J}_{z,3}$ is sizable and AFM which is largely due to the enhanced AFM coupling $\mathcal{J}^{e_{g}\textrm{-}e_{g}}_{z,3}$. The total sum of magnetic interactions in HT-phase is AFM being consistent with the calculated total energy results shown in Figure~\ref{Figure 1}; see the red diamonds which are mostly negative values\cite{comment-3}. Microscopically, the magnetic interaction and the ground state spin order of bilayer CrI$_3$ can be understood from the competition and the cooperation of FM $\mathcal{J}^{e_g\textrm{-}t_{2g}}$, AFM $\mathcal{J}^{e_g\textrm{-}e_{g}}$ and AFM $\mathcal{J}^{t_{2g}\textrm{-}t_{2g}}$ couplings.

The significantly reduced $\mathcal{J}^{e_g\textrm{-}t_{2g}}_{z,2}$ of HT phase is mainly attributed to the bond angle change as clearly seen in the analysis of maximally localized Wannier orbitals. Figure~\ref{Figure 3}a,b show that the main interaction path for $\mathcal{J}^{e_{g}\textrm{-}t_{2g}}_{z,2}$ in LT structure is the hopping between Cr-$e_g$ and $t_{2g}$ through the $e_{g}$-I$_p$ $\sigma$, I$_p$-I$_p$ $\pi$, and I$_p$-$t_{2g}$ $\pi$ bondings, which is clearly beyond the conventional superexchange process. This analysis also shows the reason why the inter-layer magnetic interaction is small ($\sim$0.1 meV); Due to the two successive hopping required, the magnetic interaction of this vdW 2D material is much weaker than the usual superexchange scale. One can further understand why this interaction $\mathcal{J}^{e_g\textrm{-}t_{2g}}_{z,2}$ is reduced in HT phase. The calculated maximally localized Wannier functions for HT phase are presented in Figure~\ref{Figure 3}c. The second neighbor $\mathcal{J}^{e_{g}\textrm{-}t_{2g}}_{z,2}$ is reduced owing to the bonding angle enlarged from 106$^{\circ}$ (LT phase) to 136$^{\circ}$ (HT phase) which leads to the weaker I$_p$-I$_p$ $\pi$ hopping (see Figure~\ref{Figure 3}c). This effect gives rise to $J^{e_{g}\textrm{-}t_{2g}}_{z,2}\sim$ 0.02 meV for HT structure which is $\sim$22\% of the LT value \cite{comment-2}.


To summarize, we investigated the magnetic interactions of bilayer CrI$_3$ from two different points of views; namely, energetics and magnetic response. Both approaches point to the same conclusion that the inter-layer AFM order could not be stabilized in the LT-structure bilayer. Further, by analyzing the orbital resolved magnetic couplings, we found that the second-neighbor $e_g$-$t_{2g}$ interaction plays the key role in stabilizing the FM order. This interaction can effectively be suppressed and become comparable with AFM $e_g$-$e_g$ interactions in HT stacking, whereby the inter-layer AFM order is stabilized. Our results provide unique information and insight to understand the magnetism of bilayer CrI$_3$ paving the way to utilize it for applications.

{{\it Note:}
	In finalizing our work, a related study \cite{sivadas_stacking-dependent_2018} is posted which contains the qualitative discussion of staking-dependent magnetism based on the calculated total energies.}


This work was supported by Basic Science Research Program through the National Research Foundation of Korea (NRF) funded by the Ministry of Education (2018R1A2B2005204) and Creative Materials Discovery Program through the NRF funded by Ministry of Science and ICT (2018M3D1A1058754). The computing resource was partly supported by the Computing System for
Research in Kyushu University.


\begin{thebibliography}{75}
\bibitem [{com( one)}]{comment-1}%
	\BibitemOpen
	\href@noop {} {} (\bibinfo {year} {The largest value of $\mathcal{J}_{z,2}$
		is partly attributed to the larger numbers of pairs. While $J_{z,1}\approx
		J_{z,2}$ per bonding (the difference is $\sim$0.01 meV), the number of second
		neighbors is nine and that of first neighbor is one.})\BibitemShut {NoStop}%
\bibitem [{com(ents)}]{comment-3}%
	\BibitemOpen
	\href@noop {} {} (\bibinfo {year} {It should be noted that the calculated
		total energy of AFM HT structure is greater than that of FM LT structure by
		0.2--6.6 meV/f.u. in our calculations. While the energy barrier between these
		two phases might be overcome in a certain stage of experimental processes, it
		still remains unclear if the HT structure was indeed realized in the previous
		experiments.})\BibitemShut {NoStop}%
\bibitem [{com(vely)}]{comment-2}%
	\BibitemOpen
	\href@noop {} {} (\bibinfo {year} {The result of Figure~2 also reflects the
		number of second neighbors which are 4 and 9 in HT and LT structure phases,
		respectively.})\BibitemShut {NoStop}%
\end{thebibliography}
\end{document}